\documentstyle[12pt]{article}
\begin{document}
\hoffset=-1.4cm
\voffset=0.4cm
\hsize=166truemm
\tolerance=5000
\def \lp{\scriptscriptstyle +}
\def \lm{\scriptscriptstyle -}
\def\sss{\rm_S}
\baselineskip=18 pt plus 2pt minus 1pt
\itemsep 3pt
\parskip  5pt plus 1pt
%
\begin{flushright}
\tabcolsep=0.0mm
\begin{tabular}{l}
Nucl.~Instr.~Meth.~{\bf A324} (1993) 535\\
CBPF NF-013-92\\
UMS/HEP/92-019\\
FERMILAB-Pub-92-137-E\\
\end{tabular}
\end{flushright}
\bigskip 
\centerline{\bf  The E791 Parallel Architecture Data Acquisition System}
\vskip 10pt
\renewcommand{\thefootnote}{\fnsymbol{footnote}}
\centerline{S.\ Amato, J.\ R.\ T.\ de Mello
Neto\footnote{Now at the Universidade Estadual do Rio de Janeiro,
 RJ, Brasil.}, and J.\ de Miranda}
\centerline{Centro Brasileiro de Pesquisas F\'{\i}sicas}
\centerline{ Rio de Janeiro \ Brasil}
\vskip 8pt
\centerline{C.\ James}
\centerline{Fermilab, Batavia, IL 60510 \ USA}
\vskip 8pt
\centerline{D.\ J.\ Summers}
\centerline{Department of Physics and Astronomy}
\centerline{University of Mississippi, Oxford, MS 38677 \ USA}
\vskip 8pt
\centerline{S.\  B.\  Bracker}
\centerline{317 Belsize Drive}
\centerline{Toronto, Ontario M4S1M7  \ Canada}
\vskip 25pt
\centerline{\bf Abstract}
\vskip 1pt
 
   To collect data for the study of charm particle decays,
we built a high speed data acquisition system for use with the E791 
magnetic spectrometer at Fermilab.  The DA system read out 24{\thinspace}000 
channels 
in 50 {$\mu$}s.  Events were accepted at the rate of 9{\thinspace}000 
per second. 
Eight large FIFOs were used to buffer event segments, which were then 
compressed and formatted by 54 processors housed in 6 VME crates.  Data was 
written continuously to 42 Exabyte tape drives at the rate of 9.6 Mb/s.
During the 1991 fixed target run at Fermilab, 20 billion physics events
were recorded on 24{\thinspace}000 8 mm tapes; this 50 Tb (Terabyte)
data set is now being analyzed.
\vfill
\eject
\leftline{\bf 1. Introduction}
\vskip 3pt
 
   Experiment 791, {\it Continued Study of Heavy Flavors}, located in
Fermilab's Proton-East experimental area, examines the properties of
short lived particles containing a charm quark.  Events involving charm quarks
are rare and difficult to recognize in real time.  The experiment's strategy
was to impose only loose constraints when recording data, and select the
events of interest offline when time and computing resources are more
available.  Therefore the DA system must collect and record data very quickly.
 
   The Fermilab Tevatron delivered beam during a 23 second spill, with a 34
second interspill period, so that the experiment generated data for 23
seconds approximately every minute. The data consists of discrete packets known
as events, each of which contains particle tracking information and calorimetry
for one interaction.  The E769 data acquisition system used previously for this
detector [1] was able to read data at 1400 kb/s during the beam spill, and
record data at 625 kb/s during both the spill and interspill; the digitizing
time per event was 840 $\mu$s.  The physics goals of E791 called for recording
at least 10 times the events collected by E769, in about the same amount of
beam-time.  The detector's digitizing and readout time had to be reduced by at
least a factor of 10; a 50 $\mu$s dead time per event was achieved by
replacing almost all the front-end digitizers with faster systems.
Events arrived at the DA
system at an average rate of 26 Mb/s during the beam spill, and were recorded
at more than 9 Mb/s during both the spill and interspill using 42 Exabyte 8200
tape drives [2].
 
   The following section will discuss the overall architecture and the hardware
components in more detail.  Following that are sections on the software used
in the DA processors, and a discussion of performance and possible upgrades.
 
\vskip 12pt
\leftline{\bf 2. Architecture and Hardware}
\vskip 1pt
 
   A schematic of the E791 DA system is shown in Fig. 1.  Events were digitized
in a variety of front-end systems and delivered into Event FIFO Buffers (EFB)
along eight parallel data paths.  The buffers stored 80 Mb of data apiece,
enough to allow the rest of the DA system to be active during both the spill and
interspill.  Care was taken to ensure that each data path carried about the
same amount of data.  Data are distributed through Event Buffer Interfaces
(EBI) to processors housed in six VME crates.  The processors (CPU) read
event segments from the buffers, compressed them into formatted events, and
recorded them on tape through a SCSI magnetic tape controller (MTC).
 
   The DA system is {\it parallel} in several respects.  Data arrives along
parallel data paths.  Processors act in parallel to prepare data for logging.
Many parallel tape drives record data concurrently.
 
\vskip 12pt
\leftline{\bf 3. Front Ends}
\vskip 1pt
 
   The E791 detector contained silicon microstrip detectors, drift
chambers, and proportional wire chambers for tracking charged
particles.  Calorimeters based on scintillators and phototubes measured
particle energies.  Gas \v Cerenkov detectors performed particle
identification, and plastic scintillators were used for muon identification.
The detector elements were digitized by various electronics systems, which
were in turn managed by front-end controllers which delivered data to the
DA system.  The front-end hardware is summarized in Table 1.
 
  The DA system placed specific requirements on the front-end controllers.
The data paths from the controllers conformed to the EFB inputs, which were
32-bit wide RS-485 lines accompanied by a single RS-485 strobe.  Data was 
delivered at a maximum rate of 100 ns per 32-bit word.  Each event segment 
on the data paths was delimited by a leading word count, calculated and placed 
there by the data path's front-end controller.  A 4-bit event synchronization 
number was generated for each event by a scaler module and distributed to all 
front-end controllers.  The controllers accepted this number and made it a part
of each event's segments.  The DA system used the synchronization number to 
assure that all event segments presented at a given moment derived from the 
same event in the detector.  Finally, because we had 16 digitizing controllers
and only 8 data paths, each data path was shared by two front-end controllers
using simple token passing.

\vskip 12pt
\leftline{\bf 4. Event FIFO Buffers}
\vskip 1pt
 
   Each Event FIFO Buffer (EFB) [3] consisted of an I/O card, a FIFO Controller
card, five 16 Mb Memory cards, and a custom backplane, housed two per crate in
9U by 220 mm Eurocrates.  The I/O card contained the RS-485 input and output
data paths, Status and Strobe lines, and a Zilog Z80 processor with a serial
port used for testing.  The Controller card kept track of internal pointers
and counters, and managed the write, read, and memory refresh cycles.
The Memory cards
used low cost 1 Mb by 8 DRAM SIMMs.  In E791, the EFBs received data in bursts
of up to 40 Mb/s and delivered data at several Mb/s concurrently.
 
   The data was pushed into the EFB's through a 32-bit wide RS-485 data port,
controlled by a strobe line driven by the attached front-end controller.
Each longword of data delivered by a front-end controller was accompanied by
the strobe which latched the data in the EFB and updated the EFB's internal
pointers.  The output side of the EFB had a similar data port and strobe,
driven by the receiving device.  The EFB maintained 4 Status lines: Full,
Near Full, Near Empty, and Empty.  The thresholds for Near Full or Near Empty
were set by the I/O card's processor.  The Near Full LEMO outputs were used in
the E791 trigger logic to inhibit triggers whenever any EFB was in danger of
overflowing.  The Near Empty Status was used by the event building processors,
and is described below.
 
\vskip 12pt
\leftline{\bf 5. Event Buffer Interface}
\vskip 1pt
 
   The EBI [4] was a VME slave module designed specifically for the E791 DA
system.  Its job was to strobe 32-bit longwords out of an EFB and make them
available to VME-based CPUs used to process events.  Figure 2 details the
connections between a single EFB and its EBIs.  Each VME crate held one EBI
for every EFB in the system, so that every CPU had access to the output
data path from every buffer.  The EFB status lines were also bussed to the
EBIs, so that the CPUs could determine how much data was available in the
buffers.  At any moment in time, only one CPU is granted control of a
particular EFB.  When a CPU in one crate is finished reading data from an EFB,
it passes control of the buffer to the next crate through a token passing
mechanism built into the EBIs.
 
   The EBI was a simple module with a few basic operations : (a) read a data
word from the EFB and strobe the next word onto the output path, (b) read the
EFB status, (c) check for the buffer control token, (d) pass the buffer control
token to the next EBI, and (e) set or clear the buffer control token.
 
\vskip 12pt
\leftline{\bf 6. VME CPUs}
\vskip 1pt
 
  The assembling of events was performed by VME based CPUs [5].  They contained
a 16 Mhz Motorola 68020 processor, a 68881 coprocessor, and 2 Mb of memory, and
were able to perform VME master single--word transfers at 2 Mb/s.  There were
8 Event Handler CPUs in each VME crate, plus one Boss CPU.  An Absoft Fortran 
compiler was available for the CPUs, and most of the E791 DA code was written 
in Fortran, except for a few time-critical subroutines which were written in 
68020 Assembler.
 
\vskip 12pt
\leftline{\bf 7. The VAX-11/780}
\vskip 1pt
 
   The VAX-11/780 was used to download and start the VME system; the DA system
operator's console and status displays were also connected to the VAX.  A low
speed link between the VAX and VME was provided by a DR11-W on the VAX Unibus,
a QBBC [5] branch bus controller, and branch bus to VME interfaces (BVI) [5]
in each VME crate.
 
\vskip 12pt
\leftline{\bf 8. Magnetic Tape Controller and Drives}
\vskip 1pt
 
   Tape writing was handled by a VME to SCSI interface, the Ciprico RF3513 [6].
The tape drives used were Exabyte 8200s writing single-density, 2.3 Gigabyte 
8 mm cassettes.  As shown in Table 2, the choice of Exabyte drives was driven 
by the media costs of storing the large amount of data we expected to record.
 
   In principle, each Magnetic Tape Controller (MTC) could be connected to 7
Exabyte drives, but we found that a single SCSI bus saturated when writing 
continuously to only four drives.  We required a data rate to tape of about 
1.6 Mb/s in each VME crate, but Exabyte drives write at a speed of only 
0.24 Mb/s.  Our solution was to use 2 MTCs per VME crate, and connect them 
to 4 and 3 Exabytes, respectively.  Thus there were 7 Exabyte drives controlled
from each VME crate, for a total of 42 drives in the DA system.
 
   The MTCs stored their SCSI commands in circular command descriptor
queues. The queues for both MTCs in a VME crate were managed by themselves
and one CPU in that crate.  The command descriptors held information on the
VME address of a block of data and the length of the block.  The MTC
acted as a VME master and performed the actual transfer of a block of
complete events from an event building CPU onto a single tape.  The
tape handling software was written to ensure that all 7 Exabyte drives
on a VME crate were filling their tapes at about the same rate.
All 42 drives were loaded with tapes at the same time, the DA system started, 
and all 42 tapes filled with data at approximately the same rate.  All the
tapes became full within a few minutes of each other, and all 42 tapes
were stopped and unloaded at the same time.  During data taking, the
tapes were full when 3 hours of beam time had elapsed.

\vskip 12pt
\leftline{\bf 9. Software}
\vskip 1pt
 
   The DA software was comprised of three main programs.  At the top was VAX,
which ran in the VAX-11/780.  It accepted user commands, generated status
displays and error logs, and fetched a tiny fraction of the incoming data to
be monitored for data quality.  Next was Boss, a program that ran in one CPU
in each VME crate.  It managed the other CPUs in its crate, and controlled
the crate's magnetic tape system.  Finally was EH, the Event Handler program
which ran in several CPUs in each VME crate.  Event Handlers did most of the
real work, reading and checking event data, formatting and compressing events,
and assembling blocks of events for eventual output to tape.  The
interprocessor communication protocol used by the three programs was the same
as used by the E769 DA system [1].
 
  Operator commands were entered on a VAX terminal, transmitted to the crate
bosses by VAX, and sent to the event handlers by Boss.  Status information
was gathered from the event handlers by Boss and compiled into a crate report;
crate reports were gathered by VAX, which generated displays and report files
for the operator.
 
  All three programs consisted of a once-only initialization code and a
processing loop which ran until the program was terminated.  Specific tasks
were placed on the processing loop, rather like beads on a string.  Each time
control passed to a task, it would proceed as far as possible without
waiting for external responses, set a flag recording its present state, and pass
control to the next task on the loop.  When that task was re-entered on the
next pass of the loop, it continued where it left off, and so on until the task
was completed.  Good real-time response was maintained while avoiding entirely
the use of interrupts.
 
\vskip 12pt
\leftline{\bf 10. Event Handler Program}
\vskip 1pt
 
   The EH program had two basic states, {\it grabber} and {\it muncher}.
Only one CPU in each crate could be in the grabber state at any given time.  
The grabber's sole duty was to read event segments from the EFBs and place them
in a large internal event array, big enough to hold 200-300 events.  When the 
crate Boss noticed that a grabber's event array was becoming quite full, it 
changed that grabber to the munching state, and appointed a new grabber.  
Because the throughput of the entire system depended on efficient event 
grabbing, grabbers were free of all other obligations, and the grabbing code 
was written in assembly language.
 
  Munchers took events from their event arrays, formatted and compressed the
data, and grouped events into physical blocks suitable for output to tape.
Munching the data could take several times longer than grabbing it, so that
at any moment each crate would have one grabber and several busy munchers.
Munchers were also subject to other obligations, such as responding to
requests for status information and binning histograms requested by the
operator.
 
  In order to achieve high system throughput from these rather slow processors,
event grabbing had to be orchestrated very carefully.  At the start of data
taking, one grabber would be appointed in each crate, and one crate would be
designated number 1.  As data arrived in the EFBs, the grabber in crate 1
would extract the event segment from EFB 1 and pass that buffer's token to
crate 2.  As the grabber in crate 1 moved on to reading the second segment of
the first event from EFB 2, the grabber in crate 2 would start reading the
first segment of the second event from EFB 1.  Soon the grabbers in all six
crates would be active, each reading from a different EFB.  Because there were 
eight EFBs but only six crates with one grabber each, all the grabbers would 
be busy all the time.
 
  Normally the crate Boss would replace a grabber with a new one before the
old grabber's event array became full.  If that reassignment were delayed, the
existing grabber would simply pass tokens through to the next crate without
reading data, giving up the event to other grabbers that might be able to
handle it.  Only if all grabbers were glutted with data and no event handlers
could be recruited as new grabbers would data taking slow down.
 
  As grabbers read data from the EFBs, they checked to ensure that the event
segment word counts were reasonable and that all event segments being joined
together in an event had the same event synchronization number.  Illegal
word counts and unsynchronized events usually indicated that a front-end
readout system had failed.  To overlook such a failure would be very serious;
pieces of unrelated data could end up being joined together into a bogus
event, and the error would propagate forward for all subsequent events.  When
such failures were noted, the grabber notified its Boss, the Boss notified
the VAX, and the VAX inhibited data taking, flushed the EFBs, and instructed
the system to restart.  Synchronization errors occurred with depressing
regularity throughout the data taking, so it was fortunate that the DA system 
had the ability to recognize and respond to them quickly and automatically.
A few spills a day were thus lost.
 
Event munching consisted of compressing the TDC data from the drift chambers
(which arrived in a very inefficient format), formatting each event so that it
conformed to the E791 standard, and packing events into tape buffers for
output.  Munchers did not control tape writing however; they submitted output
requests to their Boss, who queued the necessary commands to the tape
controller, checked the status, and notified the event handler when the tape
buffer could be reused.  Each muncher had 10 tape buffers, each capable of
holding a full-sized tape record of 65532 bytes.  Although the Boss managed
all tape writing, the data itself never passed to the Boss; the MTC extracted
the data directly from the event handler's tape buffers.
 
Most of the event munching time was spent compressing TDC data to about
$2\over3$ of its original size.  Since the TDC data was a large fraction 
of the total, it was important to compress the data, to conserve tape writing 
bandwidth and minimize tape use.  In choosing readout hardware for high-rate 
experiments, it is important to evaluate the details of the data format very 
carefully (although in this instance we had no alternate choice of vendors).
 
\vskip 12pt
\leftline{\bf 11. The Boss Program}
\vskip 1pt
 
   The CPU running the Boss program controlled the scheduling of each EH 
as a grabber or muncher.  It polled the EHs on a regular basis to check the
need for rescheduling.  The main criteria to retire a grabber and select a new
one was whether the input event arrays were full or nearly full.  When the
system was heavily loaded, protection against too frequent rescheduling
was applied.
 
   Managing tape writing was also the Boss's job.  The Boss made periodic
requests to all EHs for a list of tape buffers ready for writing.  The EHs
responded by giving the boss the VME address and the length of their full
tape buffers.  The Boss used the information to construct the commands for the
MTCs.  The  Boss also selected which MTC and tape drive to send a tape buffer
to, based on how full the MTC's command queue was and how full the tape in the
drive was.  The MTCs performed the block transfer of the tape buffer from the
EH processor to the Exabyte tape drive.  When a tape buffer was written, the
MTC informed the Boss, and the Boss in turn notified the EH that the particular
tape buffer was ready for reuse.
 
   The Bosses were also responsible for gathering status information and
reports of recoverable errors and passing the information to the VAX program.
The Bosses sent occasional Request Sense commands to the drives, which returned
the number of blocks written to tape and the number of blocks rewritten (soft
write errors).  All commands sent to the Exabyte drives were returned by the
MTC with a status block, and if a drive error occurred while writing data, the
status block gave details on the error type.  Drive errors of some types were
not recoverable, and the offending drive was taken offline until the end of
the data taking run.  Likewise, any EH which did not respond to Boss commands
within a given time limit was reset and temporarily removed from the active
system.  Event processing could continue even if a few EHs or Exabyte drives
failed since there were multiple drives and EHs in each VME crate.  The
throughput of the DA system would be slightly reduced, but not stop.
 
\vskip 12pt
\leftline{\bf 12. The VAX Program}
\vskip 1pt
 
   The VAX program managed and monitored the rest of the DA system. A
schematic is shown in Fig. 3.  The DA Control Console is shown in Fig. 4,
and provided the user with general status information and a command menu.
In regular data taking the user executed a LOAD after the tapes were placed
in the drives, then a START to begin a data taking run.  Another option was to
read out the detector without sending the events to tape (START NOTAPE).
During data taking the run could be suspended for a short time (PAUSE, RESUME)
and under special circumstances the user could clear the EFBs (CLEAR\_BUFF).
The Bosses polled the tape drives for fullness of the tapes, and sent the
information to the VAX program.  When 20\% of the active drives were 95\% full,
the VAX program automatically sent the END command.  The user could also END
data taking whenever he wished.
 
   In ending data-taking runs, it was necessary to allow a smooth run down of
the system.  The VAX first inhibited the triggers to stop the flow of data
into the EFBs.  The Bosses stopped the current grabber and did not schedule
another one.  The VAX cleared any data that remained in the EFBs, but all the
events that were already in the EH input event arrays were allowed to be
written to tape.  The VAX waited until the Bosses reported that all tape writing
was complete and file marks written before informing the user that the run was
ended.  The user could not START another data taking run or execute the tape
drive UNLOAD command until this END process was complete.
 
   The EHs stored a few events for online monitoring.  During data taking, 
the VAX retrieved these events and passed them on to an event pool managed by 
VAXONLINE software [7].  The event pool was accessible by other VAX 
workstations in the local cluster, and an entirely separate set of programs 
analyzed and displayed the pool events for online monitoring of the detector.  
Typically, the rate at which events were sent to the pool for was fast enough 
for most monitoring needs.  The DA system also provided a much faster 
alternative detector monitoring method.  Monitoring a detector typically means 
making histograms (hit maps) of the detector elements.  One can look for dead 
or noisy channels.  Part of the EH munching code constructed such histograms 
upon user request.  The user specified a particular section of the detector to 
histogram using a very simple program; the program sent the request to the VAX 
DA program using a DEC Mailbox facility.  The request was distributed to the 
VME EH processors, and all the EHs in the system would accumulate all events 
for a period of about one minute.  The Bosses and ultimately the VAX summed up 
the histogram contributions from each EH, and entered the final product into 
the event pool as a special event type.  The user's program retrieved the 
histogram from the event pool and could use a variety of means to display it.  
In this way the user could get a hit map of a part of the detector with high 
statistics, 200{\thinspace}000 events or so, in a very short time.
 
   The VAX program retrieved status information from the Bosses on a regular
basis while a data taking run was in progress.  Information such as the
numbers of events processed, the fullness of the tapes, and any errors that
occurred were displayed on various monitors and on the DA Control Console.
For every data taking run, a disk file was created which held a unique run
number, the date and time the data was recorded, the number of events written
to each drive during the run, the drive's soft error rate as a percent of
blocks written, and whether the drive failed during the run.  This file of
numbers was entered automatically into an electronic database when the run was
ended.
 
\vskip 12pt
\leftline{\bf 13. Performance and Conclusions}
\vskip 1pt
 
   The DA system hardware performed well.  As mentioned earlier, the system
was tolerant of errors encountered by CPUs running the EH program and of
Exabyte drives with write errors.  While all the hardware components in the
system experienced some infant mortality in the initial testing phases, all
the components, with one exception, had very few failures in 9 months of data
taking.  The exception was the Exabyte drives, which, after 2000 hours of
operation, will often require head replacement.  System wide failures that
halted data taking were extremely rare, and recovery if they did occur was
rapid.
 
   Running in a test mode, data was pushed into the DA system from the front end
controllers at a rate exceeding real data taking.  The DA system then gave
a maximum data rate to tape of about 9.6 Mb/s, or 1.6 Mb/s through each VME
crate.  Throughput in each part of the DA system components were well matched.
The data rate into the EFBs times the length of the beam spill matched the size
of the EFBs; the grabbing speed matched the munching speed times the number of
munchers in each VME crate; the output rate from each crate matched the tape
writing speed times the number of drives per crate.  However, during real data
taking, the maximum 9.6 Mb/s throughput was usually not attained simply
because the accelerator did not deliver enough beam to create the events.
 
   In a 5 month period of data taking in 1991 and early 1992, E791 recorded
20 billion physics events on 24{\thinspace}000 8 mm tapes.  
This 50 Tb data set is 
now being analysed at parallel RISC computing facilities similar to those used 
previously in E769 [8].  The experiment's goal of 100{\thinspace}000 
reconstructed charm 
particle decays should easily be met.
 
  The parallel architecture of the E791 DA system is central to its success.
The performance of the system could be increased with more parallel front-end
controllers for faster read out, larger Event FIFO Buffers, faster CPUs with
much better I/O capability, and by upgrading the 0.24 Mb/s Exabyte 8200 drives
to double--speed, double--density Exabyte 8500 tape drives.
 
\vskip 12pt
\leftline{\bf Acknowledgements}
\vskip 1pt
 
   We thank the staffs of all the participating institutions and especially
S. Hansen, A. Baumbaugh, K. Knickerbocker, and R. Adamo and his
group, all of FNAL.  This work was supported by the 
U.\ S.\ Department of Energy (DE-AC02-76CHO3000 and DE-FG05-91ER40622)
and the Brazilian Conselho Nacional de Desenvolvimento Cient{\'i}fico e
Tecnol{\'o}gico.

\bigskip
\bigskip
\newcounter{bean}
\leftline{\bf References}
\begin{list}
{[\arabic{bean}]}{\usecounter{bean} \setlength{\rightmargin}{-1.5cm}}
 
\item C. Gay and S. Bracker, "The E769 Multiprocessor Based Data Acquisition
      System", IEEE Trans.~Nucl.~Sci.~NS-34 (1987) 870.
 
\item Exabyte Corp., 1745 38th Street, Boulder, CO 80301, USA.

\item A. E. Baumbaugh et al., "A Real Time Data Compactor (sparsifier) and
      8 Mb High Speed FIFO for HEP", IEEE Trans.~Nucl.~Sci.~NS-33 (1985) 903;
      \newline
      K. L. Knickerbocker et al., "High Speed Video Data Acquisition System
      (VDAS) for HEP", IEEE Trans.~Nucl.~Sci.~NS-34 (1986) 245.
 
\item S.~Bracker, "Specification of the E791 Event Buffer Interface", E791 
      internal document; \newline
      S.~Hansen, FNAL Physics Dept., personal communication.

\item R. Hance et al., "The ACP Branch Bus and Real Time Applications of the
      ACP Multiprocessor System", IEEE Trans.~Nucl.~Sci.~NS-34 (1987) 878.

\item Ciprico, 2955 Xenium Lane, Plymouth, Minnesota 55441, USA.
 
\item V. White et al., "The VAXONLINE Software System at Fermilab", IEEE
      Trans.~Nucl.~Sci.~NS-34 (1987) 763.

\item C. Stoughton and D. J. Summers, "Using Multiple RISC CPUs in Parallel
      to study Charm Quarks", Computers in Physics 6 (1992) 371.

\item Phillips Scientific, 305 Island Rd., Mahwah, New Jersey 07430, USA.

\item LeCroy Research, 700 Chestnut Ridge Rd., Chestnut Ridge, NY 10977, USA.

\item C.~Rush, A.~Nguyen and R.~Sidwell, Dept.~of Physics, The Ohio State 
      University, personal communication.

\item Nanometric Systems, 451 South Blvd., Oak Park, IL 60302, USA.

\item M. Bernett et al., "FASTBUS Smart Crate Controller Manual", Fermilab
      Technical Document HN96 (1992).

\item S.~Bracker, "Description of the Damn Yankee Controller (DYC)", E791 
      Internal Document; \newline
      S.~Hansen, FNAL Physics Dept., personal communication.

\item M.~Purohit, Dept. of Physics, Princeton University, "Princeton 
      Scanner/Con{\-}troller Manual", E791 Internal Document.

\item S. Hansen et al., "Fermilab Smart Crate Controller", IEEE 
      Trans.~Nucl~Sci. NS-34 (1987) 1003.
 
\end{list}
\eject
\font\sc=cmr9
\def \longw{\sc longwords}
\noindent
{\bf{Table 1.  E791 Front End Digitization Systems and Read Out Controllers.}}
\smallskip
\vskip 8pt
\renewcommand{\arraystretch}{1.2}
{\footnotesize
\centerline{
\begin{tabular}{llllll} \hline
 System             & Drift        &  \v Cerenkov,    & Silicon Micro-        
& Proportional   & CAMAC        \\ 
                   & Chamber      &  Calorimeter     & vertex Detector      
& Wire Chamber   &              \\ \hline
 Digitizer          & Phillips [9] & LeCroy 4300B     & Ohio State [11],      
& LeCroy 2731A   & LeCroy       \\
                   & 10C6 TDC     & FERA ADC [10]    & Nanometric N339P,     
& Latch          & 4448 Latch,  \\
                   &              &                  & Nanometric S710/810   
&                & 4508 PLU,    \\
                   &              &                  & [12] Latches          
&                & 2551 Scaler  \\
 Mean Dead Time     & 30 $\mu$s    & 30 $\mu$s        & 50 $\mu$s             
      & 4 $\mu$s    & 30 $\mu$s \\
 Pre-Controllers    & none         & 2 LeCroy 4301s   & 81 Princeton Scanners 
& 2 LeCroy 2738s & none         \\
 Controller         & FSCC [13]    & Damn Yankee [14] & Princeton [15]        
& Damn Yankee    & SCC [16]     \\
 No.~of Controllers & 10           & 2                & 2                     
& 1              & 1            \\
 Channels / System  & 6304         & 554              & 15896                 
& 1088           & 80           \\
 Event Size to EFB  & 480 \longw   & 160 \longw       & 110 \longw            
& 20 \longw      & 11 \longw    \\
 Event Size to Tape & 300 \longw   & 160 \longw       & 110 \longw            
& 20 \longw      & 12 \longw    \\
 On Tape Fraction   & 50\%         & 27\%             & 18\%                  
& 3\%            & 2\%          \\
\hline \end{tabular}}}
\vskip 35 mm
\vbox{\noindent
{\bf{Table 2.  A Comparison of Storage Media.}} \quad
The 8 mm, 9-track, and 3480 tape prices are from the Fermilab stockroom
catalog. The 4 mm DAT price is from the New York Times, 20 Jan.\ 1991,
page 31.  Prices do not include overhead.
}
\smallskip
\vskip 8pt
\centerline{
\begin{tabular}{lcclrr} \hline
Tape Type & Length  & Capacity & \$/  
          & \multicolumn{1}{l}{\$/} & \multicolumn{1}{l}{Tapes/}   \\
        & [m]     & [Gb]     & tape & 50 Tbytes  & 50 Tbytes \\ \hline
8 mm video &  106   & 2.3    & \$3.92       
         & \$85{\thinspace}217   &21{\thinspace}739 \quad        \\
4 mm DAT   &   60   & 1.2    & \$7.79       
        &\$324{\thinspace}583   &41{\thinspace}667  \quad      \\
IBM 3480  &  165   & 0.22   & \$4.60       
       & \$1{\thinspace}045{\thinspace}455  & 227{\thinspace}272  \quad \\
9-track   &  732   & 0.16   & \$9.31       
       &  \$2{\thinspace}909{\thinspace}375  & 312{\thinspace}500 \quad \\
\hline \end{tabular}}
\eject
\null
\vskip150mm
\includegraphics{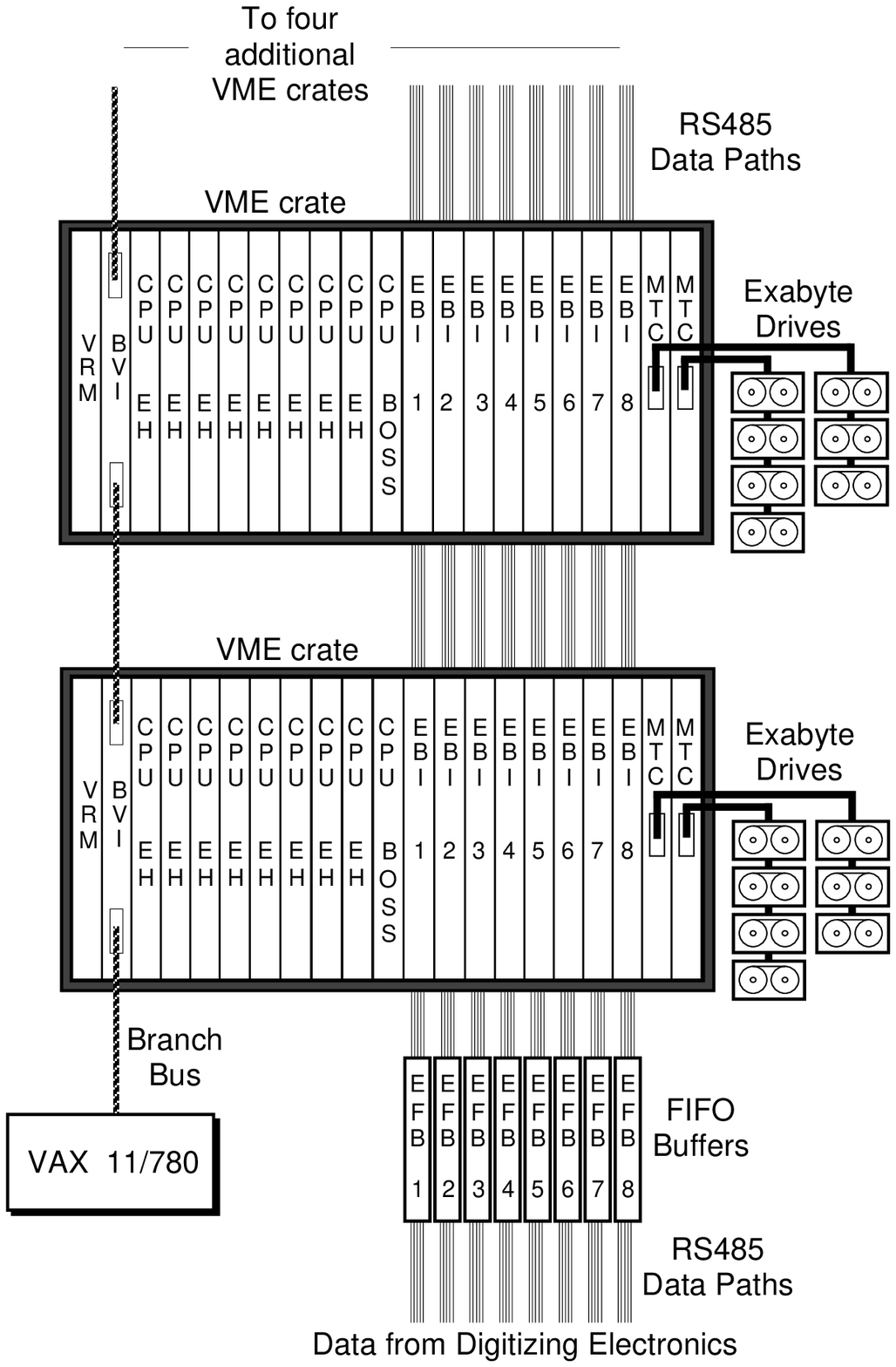}
\vfill
\vbox{\noindent
{\bf{Figure 1.}} A schematic of the VME part of the E791 DA system.  
Two complete VME crates are shown, with the Event Fifo Buffers and data paths
from the digitizers at the base.
}
\eject
\null
\vskip150mm
\includegraphics{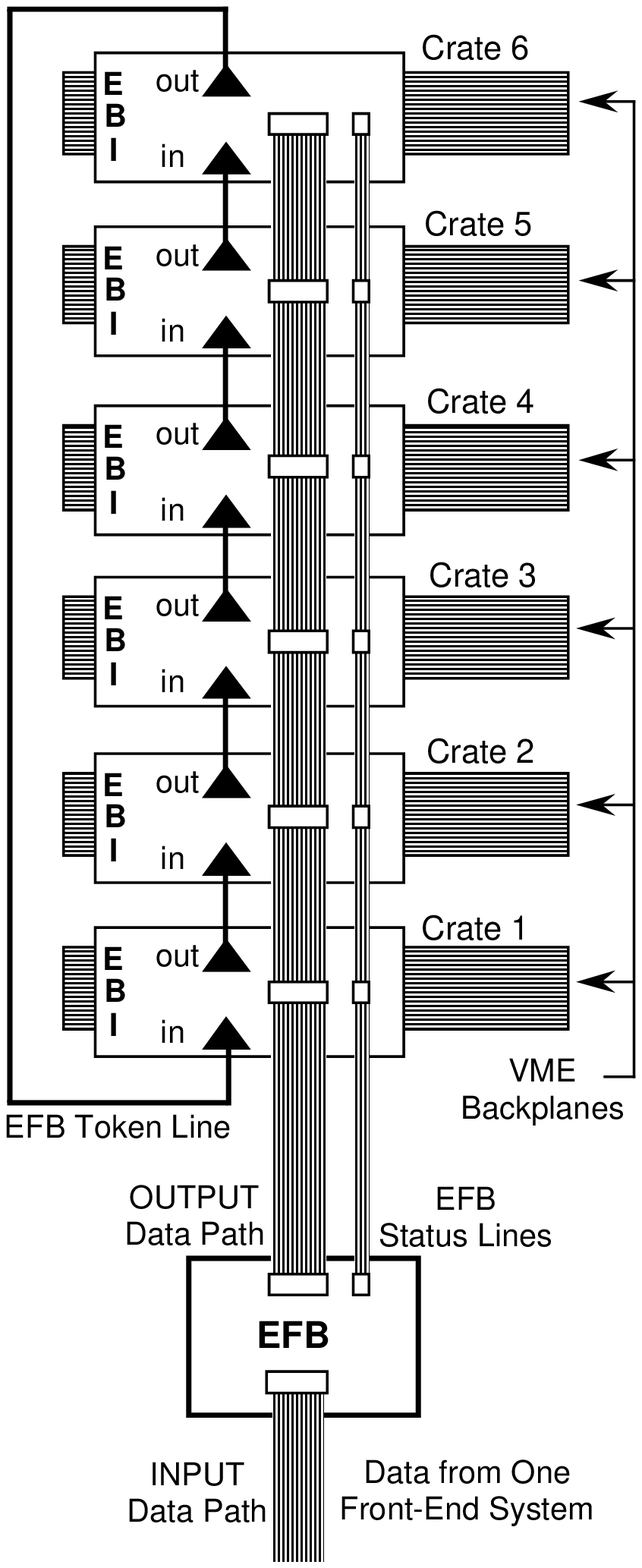}
\vfill
\vbox{\noindent
{\bf{Figure 2.}} Detail of the connections between a single EFB and the six 
EBIs attached to it.  Each EBI is in a different VME crate.  The output data 
path and the EFB status lines are bussed across all six EBIs.  The output data 
path connects to the VME backplane of each crate through the EBI.  The EBIs
share the data path by communicating along the EFB token line.
}
\eject
\null
\vskip150mm
\includegraphics{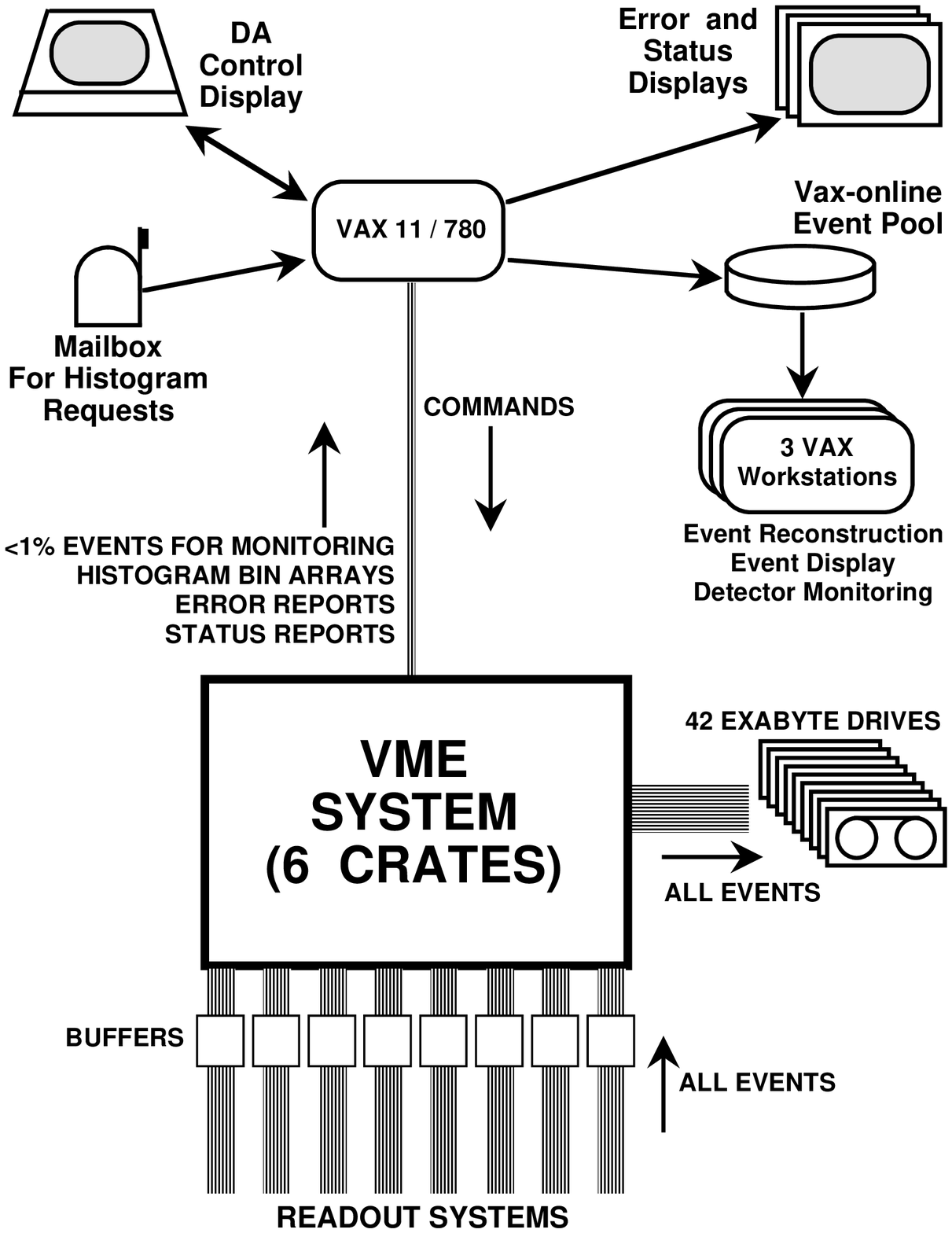}
\vfill
\vbox{\noindent
{\bf{Figure 3.}} Schematic of the entire E791 DA system.  The VAX 11/780 was
the user interface to the VME part of the system, via the DA Control Display.
The VAX part of the DA program handled the status and error displays, sent
events for monitoring to the event pool, and received histogram requests
via the mailbox.  An entirely separate set of programs picked up events from
the event pool or sent histogram requests to the mailbox.
}
\eject
\null
\vskip140mm
\includegraphics{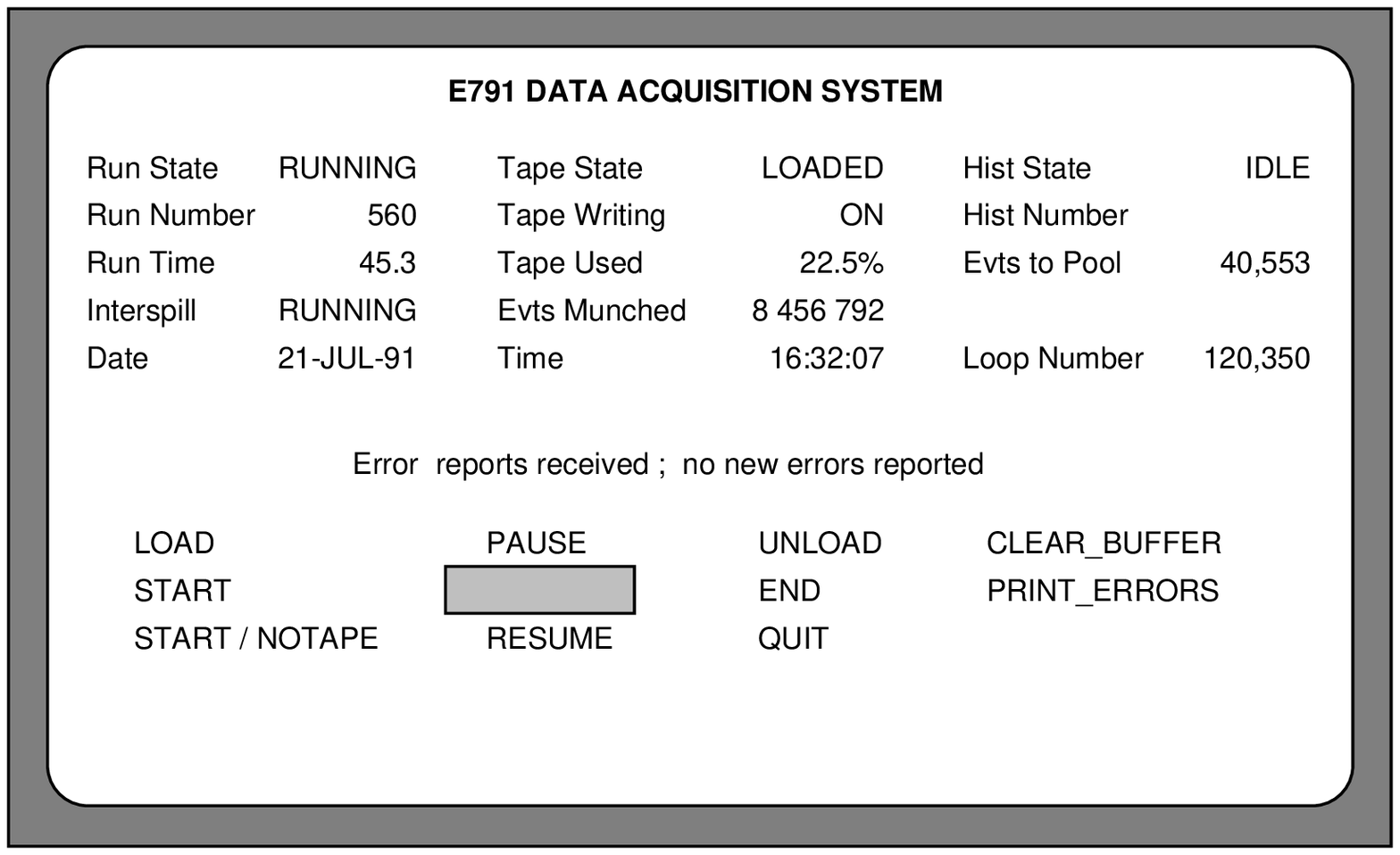}
\vfill
\vbox{\noindent
{\bf{Figure 4.}} Detail of the E791 DA Control Display.  The lower half of
the screen contained commands to the system, executed by using arrow keys to
move the shaded box over the command.  The upper half of the screen contained
contained information on the current state of the system (RUNNING or IDLE,
tapes LOADED or UNLOADED, tape writing ON or OFF), the Run Number if a 
data-taking run was in progress, and the number of events written to tape.
}
\end{document}